\documentstyle[prl,aps,amssymb,graphicx]{revtex}
\newcommand{\beq}{\begin{equation}}
\newcommand{\eneq}{\end{equation}}

\input{epsf}

\begin{document}

\tolerance 10000

\twocolumn[\hsize\textwidth\columnwidth\hsize\csname %
@twocolumnfalse\endcsname

\draft

\title{Spinon-Holon Attraction in the Supersymmetric $t-J$ Model with
        $1/r^2$-Interaction}

\author {B. A. Bernevig$^{*+}$, D. Giuliano$^\dagger$ and R. B. Laughlin$^*$}

\address{$^*$ Department of Physics, Stanford University,
        Stanford, California 94305\\
        $^\dagger$ I.N.F.M., Unit\`a di Napoli, Monte S.Angelo - via Cintia, 
        I-80126 Napoli, Italy\\
        $^{+}$ Department of Physics, Massachusetts Institute of Technology, 
        Cambridge, MA 02139}

\date{\today}
\maketitle
\widetext

\begin{abstract}
\begin{center}

\parbox{14cm}{We derive the coordinate representation of the one-spinon 
one-holon wavefunction for the supersymmetric $t-J$ model with
$1/r^2$-interaction. This result allows us to show that spinon and 
holon attract each other at short distance. The
attraction gets stronger as the size of the system is increased and,
in the thermodynamic limit, it is responsible for the square root
singularity in the hole spectral function \cite{kato}.}

\end{center}
\end{abstract}

\pacs{
\hspace{1.9cm}
}
]

\narrowtext

The low-lying excitations of one-dimensional (1-d) strongly-correlated
electron systems are not Landau's quasiparticles or quasiholes
carrying both charge and spin \cite{landau}, but rather collective
modes carrying spin-1/2, but no charge (spinons), or charge 1, but no
spin (holons) \cite{sphol}. Spinons and holons are semions,
i.e. particles with statistics half that of regular fermions
\cite{haldane001,nayak}.  In the thermodynamic limit, the physics of
1-d correlated electron systems is described by a gas of spinons and
holons \cite{haha}. The corresponding energy is additive, being the
sum of the energies of each isolated particle. Although this implies
that the spinon-holon interaction energy is irrelevant in the
thermodynamic limit, it gives no information about short-distance
dynamics.

In this letter we carefully analyze the short-distance interaction
between a spinon and a holon in an exact solution of the
supersymmetric $t-J$ model with $1/r^2$ interaction
(``Kuramoto-Yokohama'' (KY)-model'') \cite{kuramoto}.  The KY-model is
a system of electrons on a lattice with periodic boundary
conditions. Double occupancy of a site is forbidden by strong Coulomb
repulsion. Unoccupied sites are allowed and therefore holes can live
on the lattice.  Charge hopping, Coulomb, and spin-spin
antiferromagnetic terms are inversely proportional to the square of
the chord between the corresponding sites. The Hamiltonian is given by

\[
H_{\rm KY} = J ( \frac{2\pi}{N})^2 \sum_{\alpha < \beta}^N \frac{1}{ |
z_\alpha - z_\beta |^2 } \ P \biggl\{ \vec{S}_\alpha \cdot
\vec{S}_\beta
\]

\beq 
- \frac{1}{2} \sum_\sigma (c_{\alpha \sigma}^\dagger c_{\beta
\sigma} ) + \frac{1}{2} (n_\alpha + n_\beta) - \frac{1}{4} n_\alpha
n_\beta - \frac{3}{4} \biggr\} P  \;\; ,
\label{kyham}
\eneq
\noindent
where $z_\alpha = \exp(2\pi i \alpha / N)$ ($\alpha$ is a lattice
site, $N$ is the number of sites), $\vec{S}_\alpha$ is the spin
operator at site $\alpha$, $c_{\alpha \sigma}$ is the electron
operator at site $\alpha$, $n_{\alpha \sigma}$ = $c_{\alpha
\sigma}^\dagger c_{\alpha \sigma}$, and $P$ is the Gutzwiller
projector that annihilates configurations with doubly-occupied sites:
$P=\prod_\alpha (1-n_{\alpha \uparrow} n_{\alpha \downarrow} )$. 
At filling-1/2, the KY model coincides with the Haldane-Shastry (HS) model
of 1-d antiferromagnet.

\begin{figure}
\centering \includegraphics*[width=0.9\linewidth]{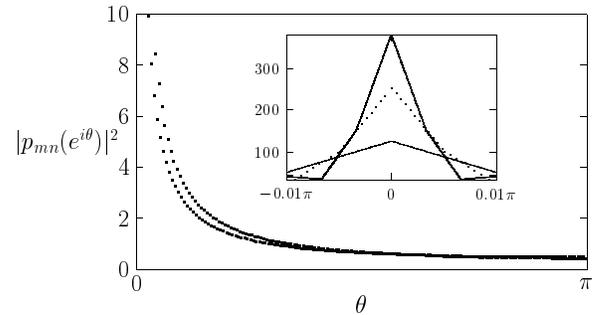}
\caption{Square of the spinon-holon wavefunction $ |p_{mn} (z)|^2$
defined by Eq.(\ref{eq20}) for $m=N/2-1$ and $N=600$. The probability
peaks up at short separation between spinon and holon, while it does
not depend on the distance at large separations. The inset shows the
function around the origin for $N=200,400, 600$. The value at the
origin diverges as $N \rightarrow \infty$.}
\label{fig1}
\end{figure}

In recent papers \cite{us1,us2}, we have worked out the real-space coordinate
representation for two-spinon wavefunctions of the HS model. 
In this letter, we extend our technique to define and work out the coordinate
representation for  one-spinon one-holon eigenfunctions and the
corresponding Schr\"odinger equation  for $H_{\rm KY}$ close to
filling-1/2, where antiholon excitations, made out of a hole
created in a holon sea, are ruled out \cite{haha,kura2}.
Spinon-holon interaction and its
nature follow from the behavior of the exact solution of this
equation. In Fig.\ref{fig1} we plot the result. The probability
does not depend on the spinon-holon separation when they are
far apart. This shows that the two particles are noninteracting at
large distances. However, at short separations, the probability is
largely enhanced, a clear evidence of a short-range attractive
interaction between the two particles. As the size of the system is
increased, the enhancement peaks up, although the interaction
contribution to the total energy decreases and becomes vanishingly
small in the thermodynamic limit  \cite{kato,kuramoto}.

The spinon-holon interaction has important consequences on the
functional form of the hole spectral density, $A_{\rm h} ( \omega,
q)$.  $A_{\rm h} ( \omega, q)$ does not exhibit Landau's quasiparticle
resonance \cite{landau}. It rather shows a sharp square-root
threshold, followed by a broad branch-cut. Broad spectra have also been
experimentally seen in ARPES experiments performed on quasi 1-d
samples \cite{zxshen}. We prove that the divergent square-root
threshold is the main effect of the attraction between a spinon and a
holon. The attraction enhances the matrix element for decay of a hole
into a spinon-holon pair, thus making the hole excitation fully
unstable to decay into a spinon and a holon. In the thermodynamic
limit, the enhancement turns into the sharp square-root threshold,
followed by the branch cut.

Let us start our analysis with some basic results from the KY-model. At
filling 1/2, the ground state of the KY-model is the same as the ground 
state of the HS-model. For even $N$, the  corresponding wavefunction 
is given by

\beq
\Psi_{\rm GS} (z_1 , \ldots , z_M ) = \prod_{i<j}^M  (z_i-z_j)^2 \prod_j^M z_j \;\;\; ,
\label{eq2}
\eneq
\noindent
where $M=N/2$ and the $\{ j\}$'s denote the locations of
$\uparrow$-spins, all the others being $\downarrow$. The corresponding
energy is given by $E_{\rm GS} = - J (\pi^2/24) (N + 5 / N )$
\cite{haldane,shastry,chia}. A $\downarrow$ spinon excitation
localized at $s$, at fixed filling, can be thought of as a
singlet sea where the spin at $s$ is constrained to be $\downarrow$. It
is described by the wavefunction

\beq
\Psi_s (z_1 , \ldots , z_M ) = \prod_j^M ( z_j - s ) z_j  
\prod_{i<j}^M (z_i - z_j )^2 \;\;\; ,
\label{eq3}
\eneq
\noindent
where now $N$ is odd and $M=(N-1)/2$. One-spinon eigenstates of
$H_{\rm KY}$ are given by spinon plane waves,

\beq
\Psi_m^{\rm sp} (z_1 , \ldots , z_M ) = \frac{1}{N} \sum_s (s^*
)^m \Psi_s (z_1 , \ldots , z_M) \;\;\; ,
\eneq
\noindent 
where $m=0,\ldots,M$ and
the total crystal momentum is $q_m^{\rm sp} = (\pi/2)N - (2 \pi / N )
(m + 1/4)$ (mod $2\pi$) The energy is given by $E_m^{\rm sp} = - J
(\pi^2 / 24 ) ( N - 1/N ) + J/2 ( 2 \pi / N)^2 m ( M-m)$. In terms of
$q_m^{\rm sp}$ the energy with repect to the ground state is $E
(q_m^{\rm sp}) = (J/2) [ (\pi/2)^2 - (q_m^{\rm sp})^2 ] $ (mod $\pi$).

One-holon states carry spin 0 and charge 1 with respect to the ground
state.  They can be constructed by removing an
electron from the center of a spinon \cite{kuramoto,haha,chia}. Unlike
in the spinon case, the Brillouin zone for one-holon states is not
halved \cite{chia}. The negative-energy part of the Brillouin zone is
spanned by the propagating one-holon wavefunctions

\beq
\Psi_n^{\rm ho} (z_1 , \ldots , z_M | h ) = h^n \prod_j^M (z_j - h ) z_j
\prod_{i<j}^M (z_i - z_j )^2 \; ,
\label{eq5}
\eneq
\noindent
where $h$ is the coordinate of the empty site and $0 \leq n \leq
(N+1)/2$. The total crystal momentum of the state in Eq.(\ref{eq5}) is
$q_n^{\rm ho} = (\pi/2)N + (2 \pi / N ) (n - 1/4)$ (mod $2\pi$). Its
total energy is given by $E_n^{\rm ho} = - J (\pi^2 / 24 ) ( N - 1/N )
+ J/2 ( 2 \pi / N)^2 n ( -(N+1)/2 + n)$. The energy with respect to
the ground state is $E^{\rm ho} (q_n^{\rm ho} ) = - (J/2) [ (\pi/2)^2
- (q_n^{\rm ho})^2 ] $ (mod $\pi$) \cite{kuramoto,haha,chia}. Here we
will not consider positive-energy one holon eigenstates \cite{chia}, 
since they are irrelevant to our analysis.

Spinons and holons do not lose their identity in a many-spinon
many-holon state. Hence, to construct one-holon one-spinon
eigenstates of $H_{\rm KY}$ we can start with states in a mixed
representation, where only the spinon is localized at $s$. Let $N$ be
even and $M=N/2-1$. The state $\Psi_s^n$ is defined as

\beq
\Psi_s^n (z_1 , \ldots z_M | h ) = h^n \prod_j^M (z_j - s ) ( z_j - h ) z_j
\prod_{i<j}^M (z_i - z_j )^2 ,
\label{eq6}
\eneq
\noindent
where $1 \leq n \leq M+2$, the spin at $s$ is constrained to be $\downarrow$
and $h$ is the coordinate of the empty site. States with a well-defined 
crystal momentum can be constructed by propagating the spinon and are
given by

\beq
\Psi_{mn} (z_1 , \ldots z_M | h ) =  \sum_s \frac{(s^*)^m}{N} 
\Psi_s^n (z_1 , \ldots z_M | h ) \;\;.
\label{eq7}
\eneq
\noindent
The total crystal momentum is $q = (\pi/2) (N-2) + q_m^{\rm sp} + 
q_n^{\rm ho}$ (mod $2 \pi$); $ q_m^{\rm sp}$ and $q_n^{\rm ho}$ are the
momenta of the single spinon and holon, respectively. One-spinon one-holon
eigenstates of the KY-Hamiltonian are linear combinations of 
the states $\Psi_{mn}$ with fixed total momentum

\beq
\Phi_{mn} = \sum_{\ell=0}^m a_\ell \Psi_{m-\ell,n-\ell}
\;\;\; {\rm if} \; m-n+1<0 \;\;\;,
\label{eqn0}
\eneq
\noindent
and
\beq
\Phi_{mn}=\sum_{\ell=0}^{M-m} a_\ell \Psi_{m+\ell,n+\ell} \;\;\; {\rm if} \; 
m-n+1 \geq 0 \;\;\;.
\label{eqen}
\eneq
\noindent
The coefficients $a_\ell$ are defined by recursion as

\beq
a_\ell = - \frac{1}{2 \ell} \sum_{k=0}^{\ell -1} a_k \;\;\; 
a_0 = 1  \;\;\; ,
\label{eq8}
\eneq
\noindent
and the corresponding eigenvalue is

\beq
E_{mn} = E_{\rm GS} +  E (q_m^{\rm sp} ) +  E (q_n^{\rm ho} ) 
 - \frac{ \pi J}{N} \frac{
| q_m^{\rm sp} - q_n^{\rm ho} |}{2} .
\label{eq9}
\eneq
\noindent
The energy of the one-spinon one-holon state relative to the ground
state is the sum of the energies of an isolated spinon and an isolated
holon plus an interaction term that is vanishingly small in
the thermodynamic limit.

In order to compute the norm of the states $\Phi_{mn}$, we employ the
recursive technique introduced in \cite{us1,us2}. $\Psi_s^n$ has the
generic form $\Psi_s^n (z_1 , \ldots , z_M | h ) = \Phi_s^n (z_1
, \ldots , z_M | h ) \Psi_{\rm GS}$, where $\Phi_s^n$ is a
symmetric polynomial in $z_1 , \ldots , z_M$, and $\Psi_{\rm GS}$ is
the wavefunction introduced in Eq.(\ref{eq2}). Define the operator
$e_1 (z_1 , \ldots , z_M ) = z_1 + \ldots + z_M$. It is
straightforward to prove that

\[
[ H_{\rm KY} , e_1 ] [ \Phi_s^n  \cdot \Psi_{\rm GS}]  =
\Psi_{\rm GS} \frac{J}{2} ( \frac{ 2 \pi}{N} )^2 \biggl\{ \biggl[ 
( M + \frac{1}{2} )  + M ( s + h )
\]

\beq
 - s^2 \frac{ \partial}{\partial s^2} 
\biggr] \Phi_s^n  +  h \frac{h}{h-s} \left[  \Phi_s^n  
- \left( \frac{s}{h} \right)^n  \Phi_h^n  
\right] \biggr\} \;\; .
\label{eq10}
\eneq
\noindent
Using Eq.(\ref{eq10}), we find the following relations between 
matrix elements of $[ H_{\rm KY} , e_1]$ between energy eigenstates, 
calculated with respect to  the inner product $\langle f | g \rangle 
= \sum_{z_1 , \ldots , z_M , h} f^* (z_1 , \ldots , z_M | h ) g  
(z_1 , \ldots , z_M | h )$ and valid in the case $m-n+1 < 0$

\beq
\frac{ \langle \Phi_{m-1,n} | e_1 | \Phi_{mn} \rangle }{
 \langle \Phi_{m-1,n} | \Phi_{m-1,n} \rangle }
= - \frac{M-m + \frac{3}{2} }{
2 (M-m+1 )} \;\; ,
\label{eq11}
\eneq
\noindent
and

\beq
\frac{ \langle \Phi_{m-1,n} | e_1 | \Phi_{mn} \rangle}{
\langle \Phi_{m,n} | \Phi_{m,n} \rangle } = 
- \frac{m}{ 2 (m - \frac{1}{2} )} \;\;\;.
\label{eq12}
\eneq
\noindent
From eqs.(\ref{eq11},\ref{eq12}) one finds by recursion that

\beq
\frac{ \langle \Phi_{mn} | \Phi_{mn} \rangle}{ \langle \Psi_{\rm GS} |
\Psi_{\rm GS} \rangle } = \frac{ \Gamma [ m + \frac{1}{2} ] \Gamma [ M - m 
+ 1]}{ N \Gamma [ m + 1 ] \Gamma [ M - m + \frac{3}{2} ]} \;\; ,
\label{eq13}
\eneq
\noindent
where $\langle \Psi_{\rm GS} | \Psi_{\rm GS} \rangle = N^{M+1} 
(2M+2 )!/2^{M+1}$ \cite{kwil}. As $m-n+1 \geq 0$, by following the same 
steps we find

\beq
 \frac{ \langle \Phi_{mn} | \Phi_{mn} \rangle}{ \langle \Psi_{\rm GS} |
\Psi_{\rm GS} \rangle } = \frac{\Gamma [ m + 1 ] \Gamma [ M - m 
+ \frac{1}{2} ]}{ N \Gamma [ m + \frac{3}{2} ] \Gamma [ M - m + 1 ]} \;\; .
\label{eq14}
\eneq
 
The state for a localized spinon at $s$ and a localized holon at
$h_0$, $\Psi_{s h_0}$ is defined as the Fourier-transform of $\Psi_s^n$
back to coordinate space

\beq
\Psi_{s h_0} = \sum_{n=1}^{M+2} h_0^{-n} \Psi_s^n \;\;\; .
\label{eq15}
\eneq
\noindent
As in \cite{us1,us2}, we define the real-space coordinate
representation for a spinon-holon pair, $s^m h_0^{-n} p_{mn} ( s / h_0
)$, as

\beq
\Psi_{s h_0 } = \sum_{n=1}^{M+2} \sum_{m=0}^M s^m h_0^{-n} p_{mn} ( s / h_0 )
\Phi_{mn} \;\; .
\label{eq16}
\eneq
\noindent
Notice that, from Eqs.(\ref{eqn0},\ref{eqen},\ref{eq16}), we find that the 
relative wavefunction, $p_{mn}$, is a polynomial in the variable 
$s/h_0$ if $m-n+1 < 0$,
and a polynomial in the variable $h_0 /s$ if $m-n+1 \geq 0$. Since $\Phi_{mn}$
is an eigenstate of the KY-Hamiltonian, we obtain

\[
( E_{mn} - E_{\rm GS})  \langle \Phi_{mn} |  \Psi_{s h_0} \rangle  = 
 \langle \Phi_{mn} | ( H_{\rm KY}- E_{\rm GS})  | \Psi_{s h_0} \rangle =
\]

\[
J ( \frac{ 2 \pi}{N} )^2 \biggl\{ \biggl[   ( M 
- s \frac{ \partial}{ \partial s} ) s \frac{ \partial}{ \partial s}
+  h_0 \frac{ \partial }{ \partial h_0} ( 1 + \frac{N}{2} + h_0 \frac{ \partial}{
\partial h_0} )
\]

\[
 + \frac{1}{2} \left( \frac{ h_0 + s}{ h_0 - s} \right) \left(
s\frac{ \partial}{ \partial s} + h_0 \frac{ \partial }{\partial h_0} + 1 
\right) \biggr]   \langle \Phi_{mn} |  \Psi_{s h_0} \rangle
\]

\beq
+ \frac{h_0}{s - h_0} \left( \frac{s}{h_0} \right)^\nu  
\langle \Phi_{mn} |  \Psi_{h_0 h_0} \rangle
\biggr\} \;\;\; ,
\label{eq17}
\eneq
\noindent
where $\nu = M$ if $m-n+1 < 0$, $\nu = 0 $ otherwise.

For $m-n+1 <0$, Eq.(\ref{eq17}) turns into the following equation of motion 
for $p_{mn}$

\beq
[ 2 \frac{d}{d z} - \frac{1}{ (1-z)} ] p_{mn} (z) + \frac{z^{M-m-1}}{ (1-z) }
 p_{mn} (1) = 0 
\label{eq18}
\eneq
\noindent
while, if $m-n+1 \geq 0$, it becomes

\beq
\left[ 2 \frac{d}{d ( \frac{1}{z} )} - 
\frac{1}{ ( 1 - \frac{1}{z} )} \right] p_{mn} ( z ) 
+ \frac{ ( \frac{1}{z} )^m }{  ( 1 - \frac{1}{z} ) } p_{mn} ( 1 ) = 0 . 
\label{eq19} 
\eneq
\noindent
Eqs.(\ref{eq18},\ref{eq19}) are Dirac-like first order differential
equations.  They contain a short-range interaction potential that
diverges at short distances as the first power of the separation
between spinon and holon. Its effects can be analyzed by studying the
corresponding exact solutions.  The solution to Eq.(\ref{eq18}) is a
polynomial in $z$

\beq
p_{mn} ( z ) = \sum_{k = 0}^{M-m-1} \frac{ \Gamma [ k + \frac{1}{2} ] }{ 
\Gamma [ \frac{1}{2} ] \Gamma [ k + 1 ] } z^k \;\;\; ,
\label{eq20}
\eneq
\noindent
while the solution to Eq.(\ref{eq19}) is a polynomial in $1/z$

\beq
p_{mn}^{'} ( z ) = \sum_{k = 0}^{m} \frac{ \Gamma [ k + \frac{1}{2} ] }{ 
\Gamma [ \frac{1}{2} ] \Gamma [ k + 1 ] } (\frac{1}{z})^k \;\;\; .
\label{eq21}
\eneq
\noindent
In Fig.\ref{fig1} we plot $|p_{mn} ( e^{i \theta })|^2$
vs. $\theta$. The sharp maximum at small spinon-holon separation is
the main effect of the strong attractive potential in 
Eqs.(\ref{eq18},\ref{eq19}). It is worth stressing that 
the interaction between a spinon and a holon has, for large $N$, exactly 
the same shape as the interaction between two spinons \cite{us1,us2}, although
the differential equation for the two-spinon wavefunction is second
order, while Eqs.(\ref{eq20},\ref{eq21}) are first order.

To rigourously prove that the spinon-holon attraction generates the
square root singularity followed by a branch cut in the hole spectral
function, $A_{\rm h} (q,\omega)$ \cite{kato}, we now calculate the
contribution to $A_{\rm h} (q,\omega)$ from one-spinon one-holon
states. Since $H_{\rm KY}$ acts on Gutzwiller-projected states, matrix
elements of $H_{\rm KY}$ between states with at least a
doubly-occupied site are zero. Accordingly, at half filling $A_{\rm h}
(\omega , q)$ takes contributions only from hole states propagating
forward. It is given by

\[
A_{\rm h} (q , \omega ) = \Im m  \biggl\{ \sum_X \frac{ 
| \langle X | \sum_{h_0} (h_0)^{-k}
 c_{h_0 \uparrow} | \Psi_{\rm GS} \rangle |^2 }{ \pi N 
\langle X | X \rangle \langle \Psi_{\rm GS} | \Psi_{\rm GS} \rangle }
\]

\beq
\times 
\frac{1}{ \omega + i \eta - (E_X - E_{\rm GS} ) } \biggr\} \;\;\; ,
\label{eq22}
\eneq
\noindent
where $|X \rangle$ is an exact eigenstate of $H_{\rm KY}$, $E_X$ is
its energy and $q = 2 \pi k / N$. In order to calculate the
contribution to $A_{\rm h} (q, \omega )$ from one-spinon one-holon states, 
$A_{\rm h}^{\rm sp\; ho}  (q, \omega )$ we
sum over $|X \rangle = |\Phi_{mn} \rangle$. From Eq.(\ref{eq15}) we obtain

\[
A_{\rm h}^{\rm sp\; ho} ( \omega , q)  =
\Im m \frac{1}{\pi}  \biggl\{ \sum_{l=2}^{M+2 } \sum_{m=0}^{l-2} 
\frac{ \delta_{k-m+l} p^2_{ml} ( 1 )}{ \omega + i \eta - (E_{mn} - E_{\rm GS} )
}  +
\]

\beq
 \sum_{l=1}^{M+2 } \sum_{m=l-1}^{M} \frac{ \delta_{k-m+l}
( p^{'}_{ml} )^2 ( 1 ) }{  \omega + i \eta - (E_{mn} - E_{\rm GS} ) }
 \biggr\} \frac{ \langle \Phi_{ml} | \Phi_{ml} \rangle}{ 
\langle \Psi_{\rm GS} | \Psi_{\rm GS} \rangle } .
\label{eq30}
\eneq
\noindent
This proves that the contribution to the hole spectral density from
one-spinon one-holon states is fully determined by the probability
enhancement due to the short-range spinon-holon attraction.

The thermodynamic limit of Eq.(\ref{eq30}) is defined as $M \rightarrow 
\infty$, with $m/M$ and $n/M$ constant. By using the identity

\beq
\sum_{k=0}^L \frac{ \Gamma [ k + \frac{1}{2} ]}{ \Gamma [ \frac{1}{2} ]
\Gamma [ k + 1 ] } = 2 \frac{ \Gamma [ L + \frac{3}{2} ]}{ 
\Gamma [ \frac{1}{2} ] \Gamma [ L + 1 ] } \;\;\; ,
\eneq
\noindent
and by approximating the gamma functions with Stirling's formula, the
thermodynamic limit of Eq.(\ref{eq30}) can be written as an integral
over the one-spinon one-holon Brillouin zone

\[
A_{\rm h}^{\rm sp\; ho} ( \omega , q )  = 
 2 \Im m  \int_0^\pi \frac{ d q_{\rm ho}}{ \pi}  \biggl\{ 
\int_0^{q_{\rm ho}}  \frac{ d q_{\rm sp}}{ \pi} 
\sqrt{ \frac{ \pi-q_{\rm sp}}{ q_{\rm sp}} }
 +
\]

\beq
\int_{q_{\rm ho}}^\pi  \frac{ d q_{\rm sp}}{ \pi}  
\sqrt{ \frac{ q_{\rm sp}}{ \pi - q_{\rm sp}}}
 \biggr\} \frac{  \delta ( q - q_{\rm sp} + q_{\rm ho} )}{ \omega -  \mu + i \eta
-E ( q_{\rm sp} , q_{\rm ho} )}  \; .
\label{eq31}
\eneq
\noindent
(In Eq.(\ref{eq31}) we have added the chemical potential $\mu=J \pi^2
 /4$, in order to fix the filling at $(N-1)/2N$. $q_{\rm sp}$ and
 $q_{\rm ho}$ are the spinon and holon momenta.) Eq.(\ref{eq31}) has been
first worked out in \cite{kato}. $A_{\rm h}^{\rm sp\; ho} ( \omega , q )$
is $\neq 0$ for $0 \leq q \leq \pi$. In this region of values of $q$, 
integration of Eq.(\ref{eq31}) is straightforward. It provides:

\[
A_{\rm h}^{\rm sp\; ho} ( \omega , q  ) =
\frac{1}{  \pi^2 J q}\sqrt{ \frac{ J [ q + \frac{\pi}{2} ]^2
- \omega }{ \omega - J [ q - \frac{\pi}{2}]^2} }
\times
\]

\beq 
\Theta \left[ \omega - J [ q - \frac{\pi}{2}]^2 \right] 
\Theta \left[ J [ \frac{\pi^2}{4} + q ( \pi - q ) ] - \omega \right] \; .
\label{eq32}
\eneq
\noindent

From Eq.(\ref{eq32}) we see that, in the thermodynamic limit, the
short-distance probability enhancement in the spinon-holon
wavefunction becomes the square-root singularity in $A_{\rm h}^{\rm sp\; ho} 
(\omega, q)$ at the threshold energy for creation of a spinon-holon pair,
followed by the broad branch cut. Spinon-holon attraction enhances the
matrix element for decay of a hole into a spinon-holon pair. The
physical consequence is the instability of the hole, which is no
longer a legitimate excitation of the system.

In conclusion, we have studied the spinon-holon interaction in an
exact solution of the $t-J$ model with $1/r^2$ interaction.  A spinon
and a holon interact by means of a short-range attraction. Although
unable to bind the two particles, it generates a probability
enhancement as they are close to each other. It corresponds to an
enhancement in the matrix element for decay of a hole into a
spinon-holon pair, which makes the hole excitation unstable.  Hence,
our result shows that spinon-holon attraction is what makes Landau's
Fermi liquid theory break down in 1-d strongly correlated electron
systems and, consequently, makes the quasiparticle resonance disappear
\cite{luttinger}.

We acknowkedge interesting discussions with A. Tagliacozzo,
G. Santoro, E. Tosatti, D. Santiago and E. Hohlfeld.  This work was
supported by the National Science Foundation under grant
No. DMR-9813899. Additional support was provided by the
U.S. Department of Energy under contract No. DE-AC03-76SF00515.


\begin{thebibliography}{99}


\bibitem{kato} Y. Kato, Phys. Rev. Lett. {\bf 81}, 5402 (1998).

\bibitem{landau} L. D. Landau, Sov. Phys. JETP {\bf 3}, 920(1957);
                 Sov. Phys. JETP {\bf 5}, 101(1957) Sov. Phys. JETP 
                 {\bf 8}, 70(1958).

\bibitem{sphol} P. A. Bares and G. Blatter, Phys. Rev. Lett. 
                {\bf 64} 2567 (1990).

\bibitem{haldane001} F. D. M. Haldane, Phys. Rev. Lett. {\bf 66}, 1529 (1991).

\bibitem{nayak} C. Nayak and F. Wilczek, Phys. Rev. Lett. {\bf 73},
               2740 (1994).

\bibitem{haha} Z. N. C. Ha and F. D. M. Haldane,  Phys. Rev. Lett. {\bf 73},
        2887 (1994).

\bibitem{kuramoto} Y. Kuramoto and M. Yokohama, Phys. Rev. Lett. {\bf 67},
         1338 (1991).


\bibitem{us1} B. A. Bernevig, D. Giuliano and R. B. Laughlin,  Phys. Rev. 
        Lett. {\bf 86}, 3392 (2001).

\bibitem{us2} B. A. Bernevig, D. Giuliano and R. B. Laughlin, 
        cond-mat/0011270, to be published in Phis. Rev.{\bf B}. 


\bibitem{kura2} M. Arikawa, Y. Saiga and Y. Kuramoto,  Phys. Rev. 
        Lett. {\bf 86}, 3096 (2001).


\bibitem{zxshen} C. Kim {\it et al.}, Phys. Rev. {\bf B 56}, 15589 (1997).

\bibitem{haldane} F.D.M.Haldane,Phys.Rev.Lett.{\bf 60},
        635 (1988).

\bibitem{shastry} B. S. Shastry, Phys. Rev. Lett. {\bf 60},
        639 (1988).

\bibitem{chia} R. B. Laughlin et al. {\it Field Theory for
        Low-Dimensional Systems}, ed. G. Morandi et al (Springer Series in 
        Solid State Sciences, Vol.131, May 2000).



\bibitem{kwil} K. G. Wilson, Jour. Mat. Phys. {\bf 3}, 1040 (1962). 


\bibitem{luttinger} J. M. Luttinger, J. Math. Phys. {\bf 4}, 1154 (1963).


\end{thebibliography}
\end{document}